# Carrier transport and electrical bandgaps in epitaxial CrN layers

*Duc V. Dinh,\* Jens Herfort, Andreas Fiedler, and Oliver Brandt*

Dr. Duc V. Dinh, Dr. Jens Herfort, Dr. Oliver Brandt
Paul-Drude-Institut für Festkörperelektronik, Leibniz-Institut im Forschungsverbund Berlin e.V.,
Hausvogteiplatz 5–7, 10117 Berlin, Germany.
E-mail: duc.vandinh@pdi-berlin.de
Dr. Andreas Fiedler
Leibniz-Institut für Kristallzüchtung, Max-Born-Straße 2, 12489 Berlin, Germany.

Keywords: *CrN, molecular-beam epitaxy, transport properties, scattering mechanisms, electrical bandgap*

The transport properties and electrical bandgap of nominally undoped ≈75-nm-thick CrN layers simultaneously grown on AlN(0001) and AlN(11$\bar{2}$2) templates using plasma-assisted molecular beam epitaxy are investigated. The layers grown on AlN(0001) and AlN(11$\bar{2}$2) exhibit (111) and (113) surface orientations, respectively. All layers exhibit antiferromagnetism with a Néel temperature of ≈ 280 K, observed by temperature-dependent magnetic and electrical measurements. Hall-effect measurements demonstrate n-type semiconducting behavior across a wide temperature range from 4 to 920 K. At low temperatures (4 − 260 K), the data show parallel conduction channels from a metallic impurity band and the conduction band. The carrier mobility exhibits a temperature dependence consistent with a nondegenerate semiconductor, governed by ionized-impurity scattering below 400 K and phonon scattering above 400 K. An analysis of the temperature-dependent carrier density between 300 and 920 K yields two activation energies associated with intrinsic conduction: 0.15 eV (with an uncertainty of −0.02/+0.10 eV), which we attribute to the fundamental bandgap, and 0.50 eV (with an uncertainty of −0.05/0.15 eV) representing a higher energy transition.

## 1 Introduction

CrN is an emerging group-13 (or -III$_B$) transition-metal nitride semiconductor with a unique combination of mechanical,[1,2] electronic,[3–8] thermoelectric,[9–13] and magnetic properties.[14–18] The collinear antiferromagnetic ordering observed in bulk CrN below its Néel temperature ($T_{\text{Néel}} = 286$ K) arises primarily from the exchange interactions between $Cr^{3+}$ ions, which possess unpaired $3d$ electrons.[14,15] Alongside the magnetic phase transition, the structural symmetry shifts from the paramagnetic rocksalt structure ($Fm\bar{3}m$) at elevated temperatures to the antiferromagnetic orthorhombic structure ($Pnma$) below $T_{\text{Néel}}$.[14,18,19] The magnetostructural phase transition in CrN layers can be affected in various ways by strain.[7,18] In particular, a tensile strain reduces $T_{\text{Néel}}$, whereas a compressive strain increases it.[18]

The transition from an antiferromagnetic to a paramagnetic behavior, along with the associated structural changes, can lead to modifications in the electrical properties of CrN, potentially affecting carrier transport and the overall conductivity. An electronic phase transition in addition to the magnetostructural one has been reported for CrN layers with different thicknesses[7,18,20] or surface orientations.[20,21] In particular, a metal-insulator transition with decreasing thickness has been observed for CrN(001) and CrN(111) layers, with the critical thickness differing by a factor of four (12 and 50 nm, respectively).[20] Similarly, CrN(001) and CrN(111) layers of the same thickness (25 nm) grown on orthorhombic NdGaO$_3$ substrates have been found to be metallic and insulating, respectively, which the authors attributed to the effects of anisotropic strain.[21] It is noteworthy that temperature-dependent resistivity measurements for these CrN(001) and CrN(111) layers do not show the characteristic change at $T_{\text{Néel}}$, suggesting a complex relationship between thickness, surface orientation, strain, and magnetic ordering.

The bandgap of CrN remains a subject of ongoing debate, with reported values spanning from an ultranarrow 0.02 to 0.7 eV, contingent on both measurement technique and sample quality.[3–8,16,18,20,22–28] Electrical transport studies[3,4,6–8,16] and optical spectroscopy techniques[5,22,23] often provide conflicting interpretations, compounded by first-principles predictions spanning both metallic and semiconducting ground states.[18,20,24–28] While numerous electrical studies have estimated the electrical bandgap of CrN via temperature-dependent resistivity measurements up to 400 K,[3,4,6–8,16] such assessments hinge critically on the assumption that carrier mobility remains constant with temperature $T$—an assertion rarely verified in detail. Since both carrier density ($n_H$) and mobility ($\mu_H$) evolve with temperature, resistivity alone may obscure the true nature of electrical transport. To disentangle these contributions, Hall-effect measurements of $n_H(T)$ and $\mu_H(T)$ are essential.

Moreover, the electronic structure of CrN is highly sensitive to strain, which can either narrow or even collapse the bandgap under tensile strain, or widen it under compressive conditions.[7] This strain dependence likely contributes to the broad variability observed in reported gap values across the available studies.

A comprehensive understanding of the electrical properties of CrN over a wide temperature range is essential for optimizing its applications in electronic and thermoelectric devices. While high Seebeck coefficients have been observed for degenerate CrN,[9,10,12,13] unlocking its semiconducting behavior can offer greater tunability of carrier density and thermal transport. Consequently, this tunability can enhance carrier mobility while maintaining low thermal conductivity, leading to superior thermoelectric performance.[13,29] Moreover, the coexistence of magnetic ordering and semiconducting behavior makes CrN a particularly attractive material for spintronic applications, where both charge and spin transport play crucial roles. Consequently, improving the understanding and control of the electrical and electronic properties of CrN is key to fully leveraging its potential in advanced electronic, thermoelectric, and spintronic technologies.

Here, we investigate the carrier transport in ≈75-nm-thick CrN(111) and CrN(113) layers simultaneously grown using plasma-assisted molecular beam epitaxy in a temperature range between 4 and 920 K. Both layers exhibit a magnetic phase transition at $T_{\text{Néel}} \approx 280$ K, as confirmed independently by electrical and magnetic measurements. Temperature-dependent Hall-effect measurements of the layers demonstrate n-type semiconducting behavior across the entire temperature range. The electrical data between 4 and 260 K reveal the presence of parallel conduction originating from a metallic impurity and a conduction band. In this extrinsic regime, the electron density and mobility in the conduction band are extracted using a standard two-band conduction model. The electrical bandgaps of CrN are estimated from the electron density in the intrinsic regime (300 − 920 K) to be 0.15 eV (with an uncertainty of −0.02/+0.10 eV) and 0.50 eV (with an uncertainty of −0.05/0.15 eV). The electron mobility is limited by ionized-impurity (100 < T < 400 K) and phonon scattering (T > 400 K).

## 2 Results and Discussion

### 2.1 Temperature-dependent resistivity

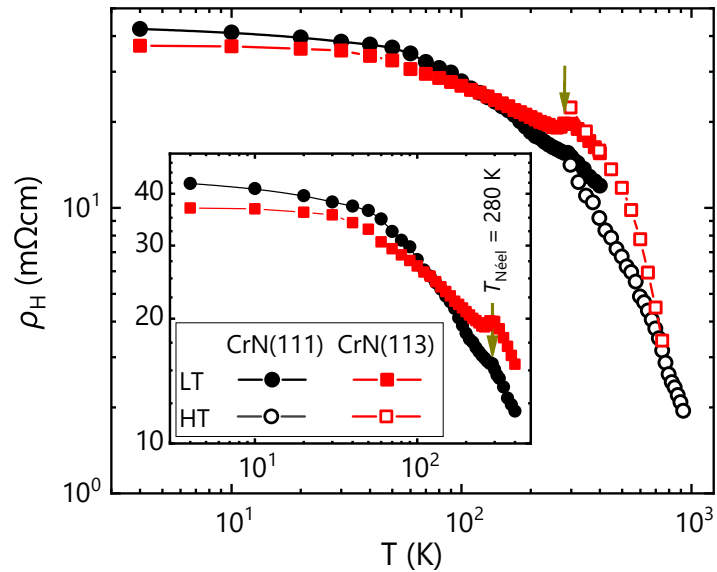

Figure 1: Temperature-dependent resistivity ($\rho_{\text{H}}$) of the CrN(111) and CrN(113) layers. Layers in sets #1 and #2 were measured in the low-temperature (LT) and high temperature (RT) range, respectively, i. e., at 4–400 K and 300–920 K. The *inset* shows their data in the LT range, clearly showing the abrupt resistivity change at a temperature at about 280 K (arrows). Note that the *inset* shares axis units and labels with the main plot.

Figure 1 presents the temperature-dependent resistivity ($\rho_{\text{H}}$) of the CrN(111) and CrN(113) layers from both sets #1 and #2. Layers with different surface orientations exhibit comparable $\rho_{\text{H}}$, with only minor deviations, as well as similar $n_H$ and $\mu_H$ (see below). All samples consistently display semiconducting behavior across the entire temperature range, as evidenced by the monotonic decrease in $\rho_H$ with increasing temperature. Hall-effect measurements indicate that electrons are the majority charge carriers, as further confirmed by the $n_H$ data shown below.

In addition, $\rho_H$ of both layers exhibits an abrupt change at $\approx$ 280 K, which is also observed in separate magnetic measurements (see Figure S10 in the supplementary material). This jump in resistivity signifies the magnetostructural phase transition at $T_{Néel} = 286$ K in accordance with the antiferromagnetic ground-state nature of bulk CrN,[14,15] and is characteristic for stoichiometric crystals with comparatively low defect densities. In fact, deviations from stoichiometry—such as an excess of Cr—can fundamentally alter the magnetic properties of CrN, thus inhibiting the phase transition.[30] Note that the magnetostructural phase transition evident in Figure 1 is not accompanied by an electronic one: the layers are semiconducting both below and above $T_{Néel}$ down to much lower temperatures.

At the lowest temperatures ($T \leq 30$ K), $\rho_H$ becomes essentially independent of temperature. The gradual transition from an activated to a metallic behavior indicates that conduction is not governed solely by the conduction band (CB) but also involves an impurity band (IB), giving rise to a two-band transport regime at low to intermediate temperatures.[31–42] Importantly, we propose that this transition between a metallic and a semiconducting behavior does not reflect an actual metal–insulator transition, but rather a shift in the dominant conduction mechanism in a heavily doped (but nondegenerate) semiconductor. This proposition motivates a more detailed investigation of carrier transport in the following subsection.

## 2.2 Carrier transport in the extrinsic regime: two-band conduction

### 2.2.1 Conduction mechanisms

Hall-effect measurements of both layers from set #1 show n-type semiconducting behavior with $n_H = 2.3 - 2.7 \times 10^{19}$ cm$^{-3}$ and $\mu_H = 12 - 19$ cm$^2$V$^{-1}$s$^{-1}$ at room temperature. To put these values into context, bulk CrN typically exhibits electron densities in the range von $5 \times 10^{20}$ cm$^{-3}$ and a mobility of about 1 cm$^2$V$^{-1}$s$^{-1}$.[43] Depending on stoichiometry, thin films may show significantly lower or higher values for $n_H$. For example, values of $2 \times 10^{19}$ and $4 \times 10^{21}$ cm$^{-3}$ were observed for sputtered CrN/MgO(001) films synthesized under N-rich and Cr-rich conditions, respectively, with values of $\mu_H$ on the order of 3 cm$^2$V$^{-1}$s$^{-1}$.[30] Mobilities as low as 0.1 cm$^2$V$^{-1}$s$^{-1}$ were measured for CrN(001) films sputter-deposited on MgO(001), and attributed to variable range hopping between individual sites in an IB.[6] On the other hand, a record high mobility of 125 cm$^2$V$^{-1}$s$^{-1}$ was reported for sputtered epitaxial CrN/Mg(001) films after thermal annealing, with an as-grown value of 1.2 cm$^2$V$^{-1}$s$^{-1}$. This remarkable increase was attributed to a reduction in the density of N vacancies in the annealed film.[43]

Figure 2 presents temperature-dependent electron density $n_H$ and electron mobility $\mu_H$ derived from Hall-effect measurements for the CrN(113) layer from set #1. Prior to interpreting this data, we need to consider the possibility of a contribution of the anomalous Hall effect (AHE) due to the antiferromagnetic nature of CrN between 4 and 286 K. In fact, the AHE has been observed in certain antiferromagnetic materials,[44–46] including Cr thin films,[44] due to mechanisms such as spin canting or symmetry breaking. However, it is very unlikely that the AHE contributes to the Hall effect in CrN (see also the further discussion of the magnetic measurements in Figure S10 in the supplementary material). Below $T_{Néel}$, CrN exhibits a collinear antiferromagnetic ordering, with the spins of adjacent atoms aligning oppositely, resulting in zero net magnetization.[14] The AHE relies on net magnetization to induce a Berry curvature in momentum space,[47,48] which is effectively absent in stoichiometric CrN such as the layers under investigation.

Returning to Figure 2, we see that for $T \leq 30$ K, both $n_H$ and $\mu_H$ remain nearly constant, consistent with metallic IB conduction. Above this temperature, $\mu_H$ increases monotonically up to 400 K. In contrast, $n_H$ increases up to 240 − 260 K, then sharply decreases near 300 K, followed by a rapid increase for $T > 300$ K. The increase in both $n_H$ and $\mu_H$ reflect activation of electrons from the IB to the CB. The nonmonotonic behavior of $n_H$ is linked to the magnetostructural phase transition occurring at $T_{Néel} \approx 280$ K. As $T$ approaches $T_{Néel}$ from below, the weakening antiferromagnetic ordering can reduce $n_H$ by altering the band structure. However, $\mu_H$ continues to increase, presumably due to the progressive thermal activation of carriers into the CB. Above $T_{Néel}$, the transition to a paramagnetic state, combined with thermal excitation of carriers from impurities states, leads to the observed increase in $n_\mathrm{H}$ up to 400 K.

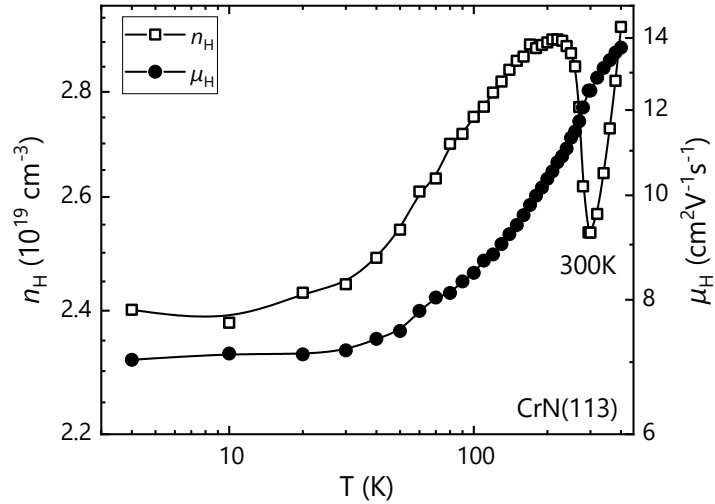

Figure 2: Temperature-dependent Hall measurements yielding $n_\mathrm{H}$ and $\mu_\mathrm{H}$, plotted on a double-logarithmic scale, of the CrN(113) layer from set #1.

The temperature dependence of $n_\mathrm{H}$ and $\mu_\mathrm{H}$ below $T_\mathrm{Néel}$ is characteristic of various semiconductors, reflecting a gradual transition from transport in the IB at low temperatures to the CB transport at higher temperatures.[32–42,49] We have also observed this behavior in our recent work on ScN.[31] Depending on impurity concentration and its Bohr radius, transport in the IB may proceed via hopping with a finite activation energy or display a metallic character as in the present case. Building on our previous experience with ScN,[31] we apply the standard two-band conduction model to the CrN layers under investigation.

### 2.2.2 Electron density and mobility in the conduction band in the extrinsic regime

To separate the contributions from the IB and CB, we apply the standard two-band conduction model[31,33] with the following expressions for the effective electron density $n_\mathrm{H}$ and mobility $\mu_\mathrm{H}$:

$$n_\mathrm{H} = \frac{(n_\mathrm{CB}\mu_\mathrm{CB} + n_\mathrm{IB}\mu_\mathrm{IB})^2}{n_\mathrm{CB}\mu_\mathrm{CB}^2 + n_\mathrm{IB}\mu_\mathrm{IB}^2}, \tag{1a}$$

$$\mu_\mathrm{H} = \frac{n_\mathrm{CB}\mu_\mathrm{CB}^2 + n_\mathrm{IB}\mu_\mathrm{IB}^2}{n_\mathrm{CB}\mu_\mathrm{CB} + n_\mathrm{IB}\mu_\mathrm{IB}}, \tag{1b}$$

where $n_\mathrm{CB}$, $\mu_\mathrm{CB}$, $n_\mathrm{IB}$ and $\mu_\mathrm{IB}$ are the electron densities and mobilities in the CB and IB, respectively. Due to the conservation of the total electron density $n_\mathrm{ext}$ in two-band conduction due to doping (the extrinsic regime), we obtain the relation

$$n_\mathrm{ext} = n_\mathrm{CB} + n_\mathrm{IB} = N_\mathrm{d} - N_\mathrm{a}, \tag{2}$$

where $N_\mathrm{d}$ and $N_\mathrm{a}$ are the densities of donors and compensating acceptors, respectively. Note that this conservation of $n_\mathrm{ext}$ does not hold for intrinsic conduction in which the carrier density is not conserved, since electron-hole pairs are generated by thermal excitation across the band gap.

To solve Equation 1, we need one additional assumption. As in our previous work on ScN,[31] we assume $\mu_\mathrm{IB}$ to be constant in the entire temperature range because of the metallic conductivity at low temperatures. Finally, for a physically meaningful positive solution, we have to set the value for $n_\mathrm{ext} > \mathrm{Max}(n_\mathrm{H})$, and the one for $\mu_\mathrm{IB} < \mathrm{Min}(\mu_\mathrm{H})$.

These assumption allow us to solve Equation 1 for the two remaining unknowns: $n_\mathrm{CB}$ and $\mu_\mathrm{CB}$.

The values of $n_\mathrm{CB}$ and $\mu_\mathrm{CB}$ extracted with this procedure are displayed along with the experimental values in Figures 3(a)–(b) for the CrN(113) layer. Clearly, electron transport is dominated by the IB at $T < 50$ K, but its influence rapidly diminishes for higher $T$. At the same time, $n_\mathrm{CB}$ increases with $T^{3/2}\exp\left(-E_\mathrm{A}/k_\mathrm{B}T\right)$

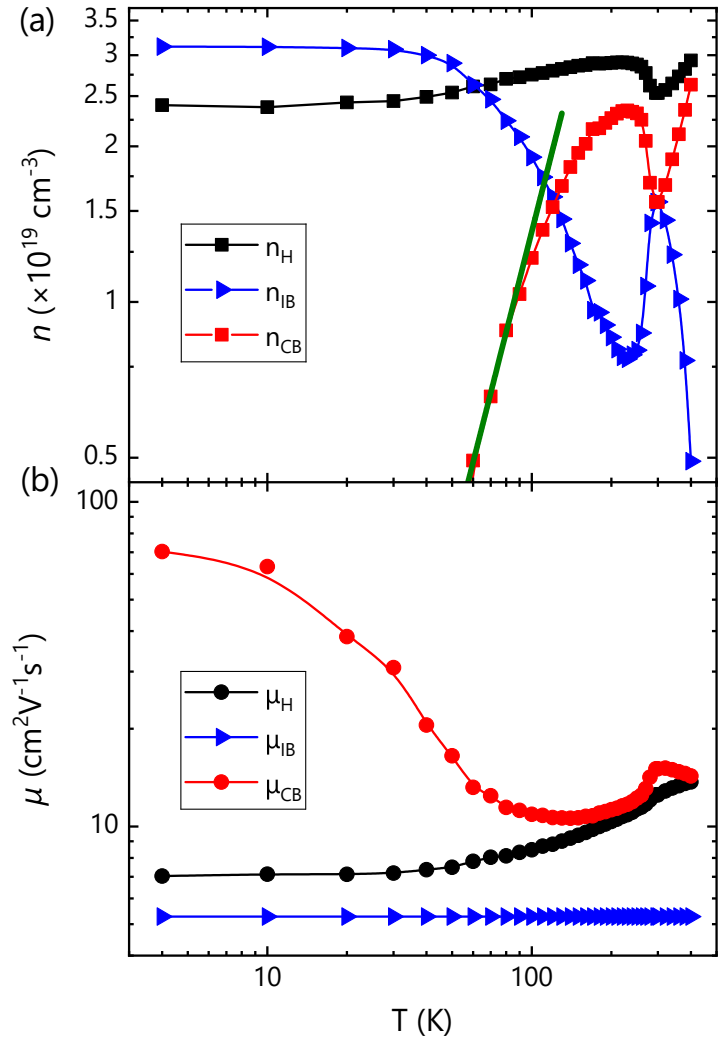

Figure 3: Double-logarithmic plots of $n_H$ (a) and $\mu_H$ (b) for the CrN(113) layer. Values of $n_{CB, IB}$ in (a) and $\mu_{CB, IB}$ in (b) are calculated using Equations 1 and 2. The line in (a) represents a fit used to extract the activation energy.

between 20 and 100 K due to activation from the IB to the CB with an activation energy $E_A$ of about 7 meV. As can be clearly seen from the comparison of the experimental data with the fit in this impurity ionization range (or freeze-out range), the slope of $n_{CB}$ tapers off at temperatures > 120 K until $n_{CB}$ reaches saturation at 200 K, and thus the impurity exhaustion range.[50] The subsequent abrupt dip is due to the magnetostructural phase transition at $T_{Néel} \approx 280$ K, as already discussed above. Finally, at temperatures > 300 K, $n_{CB}$ steeply increases with a slope much larger than in the ionization range. We attribute this increase to the onset of intrinsic conduction. The validity of the present two-band analysis is restricted to the extrinsic regime, and thus ends at about 260 K.

As for the mobility shown in Figure 3(b), we see that our analysis reveals a significantly increasing value for $\mu_{CB}$ with decreasing temperature, unlike the behavior expected for semiconductors with higher purity. The reason for this unusual behavior is the presence of the IB. At low temperatures, essentially all electrons reside in the IB, and only a very few donors are actually ionized. Consequently, ionized impurity scattering, which is usually the dominant scattering mechanism at low temperatures, becomes stronger at higher temperatures due to the activation of electrons from the IB to the CB, resulting in an exponential increase in the density of ionized donors acting as scattering centers.

The activated nature of the electron density in the CB, as well as the strong variation of the corresponding electron mobility, are characteristic of a nondegenerate semiconductor. At the same time, the IB conduction is clearly metallic. We can use these two experimental facts to get an estimate of the effective electron mass $m_e$ in the lowest CB of CrN. Nondegeneracy is equivalent to an electron density lower than the effective CB density of

states $N_{CB}$ given by

$$N_{CB} = \frac{1}{\sqrt{2}} \left( \frac{m_e k_B T}{\pi \hbar^2} \right)^{3/2},$$ (3)

where $k_B$ and $\hbar$ are the Boltzmann and the reduced Planck constant, respectively. For $N_{CB} \gg n_{CB}$ at 250 K, we need an effective mass $m_e$ (in units of the electron mass $m_0$) notably larger than 1.

At the same time, metallic impurity conduction requires an electron density higher than the critical (Mott) density for the metal-insulator transition in the lattice of impurity sites:[51]

$$n_c^{1/3} a_B = 0.26,$$ (4)

where $a_B$ is the Bohr radius of the impurity given by $a_B = (\epsilon_s/m_e) a_H$, with $a_H$ being the Bohr radius of the H atom. Taking the value of 53 for the static relative permittivity $\epsilon_s$ of CrN from Ref. 5, we arrive at a maximum value of $m_e$ of 3.5.

The range of $1 < m_e < 3.5$ obtained from these two estimates compares to a value of 4.9 obtained in Ref. 6. Although the values do not match, our rough estimate lends independent support to an exceptionally heavy electron mass in CrN.

# 3 Carrier transport in the intrinsic regime

## 3.1 Electrical bandgaps

In all previous studies, the electrical bandgap of CrN has been estimated from measurements of the resistivity that were limited to temperatures below 400 K.[3,4,6–8,16,18] This approach is inherently inaccurate for several reasons. Most importantly, the temperature range is too narrow to unambiguously discriminate between the extrinsic and the intrinsic regimes [see Figure 3(a)]. In addition, for a nondegenerate semiconductor, the mobility may be strongly temperature dependent, with the consequence that the slope of $\rho_H(T)$ does not correspond to the one of $n_H(T)$. For a reliable determination of the band gap $E_G$, it is essential to (i) measure both $n_H(T)$ and $\mu_H(T)$, and to (ii) extend the measurements to a significantly higher temperature where intrinsic conduction dominates over the extrinsic one. To this end, we utilize a dedicated high-temperature Hall setup to obtain electrical data for sample set #2 in the temperature range between 300 and 920 K.

Figure 4(a) shows the dependence of $n_H$ over the entire temperature range from 4 to 920 K for all samples under investigation. Evidently, the steep increase of $n_H$ observed in Figure 3(a) continues up to the highest temperatures observed in the present work. This observation confirms our interpretation in Sec. 2.2.2: the increase in electron density from about 300 K originates from intrinsic conduction.

A quantitative analysis of these data is now rather straightforward. As we have seen in Figure 3, the onset of intrinsic conduction occurs when extrinsic conduction is in the exhaustion range. At temperatures > 300 K, the CB is occupied by both electrons activated from donors (extrinsic ones) and electrons excited across the band gap of CrN, i.e., intrinsic ones. Since they share the same CB, they have equal mobility, and we can thus obtain the density of the latter by simply subtracting the extrinsic contribution from the total experimental density:

$$n_{int} = n_H - n_{ext}$$ (5)

Figure 4(b) presents $n_{int}(T)$ of the CrN(111) and CrN(113) layers from set #2, revealing two distinct slopes from 300 to 600 K and from 600 to 920 K. We attribute these slopes to intrinsic conduction with excitation across two distinctly different band gaps, each according to

$$n_{int}(T) = A T^{3/2} \exp(-E_G/2k_B T),$$ (6)

where $A$ is a constant. The intrinsic regime is an inherent example of two-band conduction, for which the sign and magnitude of the Hall coefficient depend on the ratio between the mobilities of electrons and holes. Experimentally, the Hall measurements attest to n-type conductivity also in the intrinsic regime. We thus assume that we can neglect the contribution of holes because their mobility is significantly lower than that of electrons, presumably because their mass is even heavier. This conclusion also justifies the use of Equation 6 valid only in the nondegenerate case, for which the effective electron mass $m_e$ has to be larger than 2 at the highest temperature of 920 K. As discussed above, also this value is still well below the one of 4.9 available in the literature.[6]

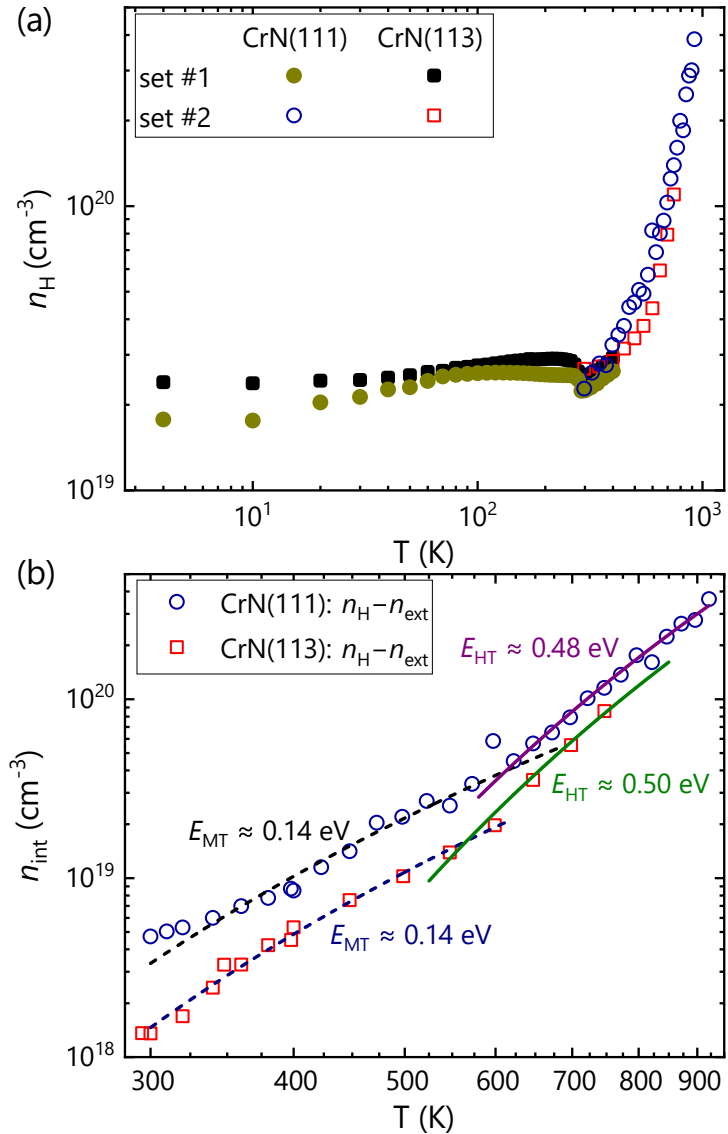

Figure 4: (a) Double-logarithmic representaion of the temperature dependent $n_H$ for the CrN(111) and CrN(113) layers from sets #1 and #2. (b) Effective intrinsic electron density $n_{int} = n_H - n_{ext}$ between 300–920 K on a double-logarithmic scale. The experimental data are fitted using Equation 6 in two temperature ranges: from 300 to 600 K (dashed line) and above 600 K (solid line).

Individual fits with Equation 6 as shown in Figure 4(b) return values of about $(0.14 \pm 0.00)$ and $(0.49 \pm 0.01)$ eV for $E_G$. Considering the various assumptions entering the two-band analysis, as well as the uncertainties in both $n_H$ and $n_{ext}$ that directly influence the slope of $n_{int}$, the overall error bar is certainly much larger, and tends to favor larger values for the band gaps. As a conservative estimate, we suggest 0.15 eV (with an uncertainty of $-0.02/+0.10$ eV) and 0.5 eV (with an uncertainty of $-0.05/0.15$ eV) as values for the actual electrical band gaps of CrN. Note that the reduction of the bandgap with increasing temperature affects the prefactor $A$ in Equation 6, but not the slope .[35]

## 3.2 Mobility and scattering mechanisms

In semiconductors as diverse as ScN,[31] Ge,[52] GaAs,[53] GaN,[54] and $\beta$-Ga$_2$O$_3$[55] the low-field scattering mechanisms have been effectively studied by comparing temperature-dependent mobility data to individual extrinsic and intrinsic scattering mechanisms combined using Matthiessen's rule.[56,57] The antiferromagnetic ordering in CrN can, in principle, influence transport properties through mechanisms such as spin-carrier coupling or spin-disorder scattering,[25–28,58–60] but since we do not observe any change of $\mu_H$ at $T_{\text{Néel}} \approx 280$ K, we can safely disregard these mechanisms.

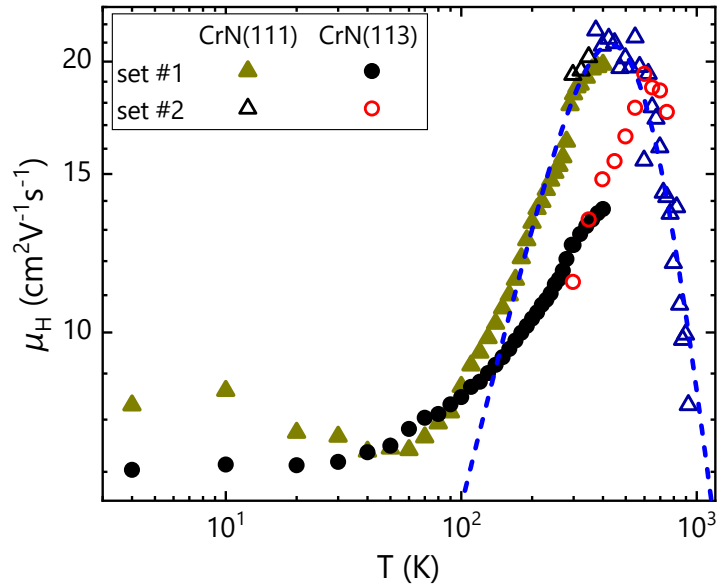

Figure 5: Double-logarithmic representation of the temperature dependent $\mu_H$ for the CrN(111) and CrN(113) layers from sets #1 and #2. The line shows a fit for $T > 200$ K by Equation 7.

Examining Figure 3(b) again, it is clear that the experimental values of $\mu_H$ are a reasonable approximation for $\mu_{\text{CB}}$ from about 200 K, where activation from the IB to the CB is essentially complete. In the intrinsic range starting at 300 K, $\mu_H$ will even closer approximate $\mu_H$, so that $\mu_H$ can be used directly to analyze the scattering mechanisms of electrons residing in the CBs of CrN. Since both $n_{\text{ext}}$ and $n_{\text{int}}$ share the same conduction bands, they should also have the same mobility, and there is no need to subtract an offset as for the electron density in Sec. 3.1.

Figure 5 shows that $\mu_H$ monotonically increases above 200 K, reaches a maximum around 400 K, and then decreases at higher temperatures. The increase from 200 to $\approx$ 400 K is characteristic of ionized-impurity scattering (or other charged defects such as dislocations,[54]) while the decrease at $T > 400$ K can be ascribed to phonon scattering, possibly involving acoustic and/or optical phonon modes.

To describe this behavior quantitatively, we consider a phenomenological expression combining these two terms: $PT^{-\alpha}$ at $T < 400$ K and $QT^{\gamma}$ at $T > 400$ K:

$$\mu_{\text{CB}}(T) \approx \mu_H(T) = [PT^{-\alpha} + QT^{\gamma}]^{-1}, \tag{7}$$

where the exponents (or slopes) $\alpha$ and $\gamma$ characterize the temperature dependence of ionized-impurity and acoustic phonon scattering, respectively, with $P$ and $Q$ being constants. The fit returns a value of $\alpha$ of about 1, which is smaller than the value of 1.5 expected for ideal nondegenerate semiconductors.[50,52–54,61] Indeed, the effective slope at high impurity and carrier densities may be smaller than this value due to screening.[50,54,61]

In contrast, the value of $\gamma$ returned by the fit is approximately 2.3, notably larger than the value of 1.5 associated with acoustic phonon scattering. However, a fit assuming polar (longitudinal) optical phonon scattering returns physically implausible values for the phonon energy, and is thus not a credible alternative interpretation. Possibly, the steep power law dependence observed here arises from a softening of the transverse optical phonon mode reported for CrN,[62] which increases the phonon population and enhances carrier-phonon scattering at elevated temperatures.[62] The softening can thus steepen the temperature dependence of $\mu_H$ beyond the conventional acoustic phonon limit.[62]

## 4 Summary and Conclusions

We have investigated the electrical properties of nominally undoped CrN layers with (111) and (113) orientations, simultaneously grown on AlN templates by plasma-assisted molecular beam epitaxy. Temperature-dependent electrical measurements from 4 to 920 K reveal semiconducting behavior for both orientations. Conductivity measurements from 4 to 260 K indicate parallel conduction channels arising from metallic impurity states and the conduction band. A two-band conduction model is used to extract the electron density and mobility in the conduction band and reveals a nondegenerate semiconductor. Analysis of the electron density data in the high-temperature range yields two characteristic activation energies corresponding to electrical bandgaps of 0.15 eV (with an uncertainty of $-0.02/+0.10$ eV), which we attribute to the fundamental bandgap, and 0.50 eV (with an uncertainty of $-0.05/0.15$ eV) representing a higher energy transition. The electron mobility is limited by ionized-impurity and phonon scattering below and above 400 K, respectively. Taking into account the metallic IB conduction, and the nondegenerate behavior up to the highest temperatures, we obtain a range for compatible values of the effective electron mass $m_e$, namely, $2 < m_e < 3.5$.

## 5 Experimental Section

Nominally undoped CrN layers are grown simultaneously on insulating AlN(0001)/Al$_2$O$_3$(0001) and AlN(11$\bar{2}$2)/Al$_2$O$_3$(10$\bar{1}$0) templates by plasma-assited molecular beam epitaxy (PAMBE), which were fabricated by metal-organic vapor phase epitaxy.[63,64] Before being loaded into the ultrahigh vacuum environment, these templates are cleaned in a HCl solution to remove the surface oxide as well as surface contaminants, and then rinsed with de-ionized water and finally blown dry with a nitrogen gun. Prior to CrN growth, the templates were outgassed for two hours at 500 °C in a load-lock chamber attached to the MBE system. The MBE growth chamber is equipped with high-temperature effusion cells to provide 4N-pure Cr and 6N-pure Al metals. A Veeco UNI-Bulb radio-frequency plasma source is used for the supply of active nitrogen (N*). 6N-pure N$_2$ gas is used as N precursor, which is further purified by a getter filter. The N* flux is calculated from the thickness of a GaN layer grown under Ga-rich conditions, and thus with a growth rate limited by the N* flux. Prior the CrN growth, a 20-nm-thick AlN layer was grown at a thermocouple temperature of 1000 °C (corresponding to a pyrometer temperature of $\approx$ 950 K) under slightly Al-rich conditions. Afterwards, the samples are cooled to 500 °C, and the CrN layers are subsequently grown at the same temperature under N*-rich conditions (Cr/N* ratio $\approx$ 0.1). These growth conditions are critical for suppressing the formation of the hexagonal Cr$_2$N phase. The thickness of the layers has been estimated to be of $\approx$ 75 nm by fitting the experimental triple-axis $2\theta - \omega$ x-ray diffraction (XRD) scans (see Figure S1 in the supplementary material).

The structural properties of the CrN/AlN layers are characterized using a high-resolution XRD system (Philips Panalytical X'Pert PRO MRD) equipped with a two-bounce hybrid monochromator Ge(220) for the CuK$_{\alpha 1}$ source ($\lambda$ = 1.540 598 Å). The layer grown on the AlN(0001) template exhibits a pure (111) surface orientation with twinned domains (see Figure S2 in the supplementary material). The layer grown on the AlN(11$\bar{2}$2) template also shows twinned domains but with a pure (113) surface orientation (see Figures S3–S4 in the supplementary material), which is similar to our recent report on the growth of ScN(113) on AlN(11$\bar{2}$2).[65] Both layers exhibit twinned domains, where the in-plane relationship between CrN(111) and AlN(0001) is [11$\bar{2}$]$_\text{CrN}$ || [1$\bar{1}$00]$_\text{AlN}$ (see Figure S2 in the supplementary material). For CrN(113) and AlN(11$\bar{2}$2), the relationship is [33$\bar{2}$]$_\text{CrN}$ || [1$\bar{1}$00]$_\text{AlN}$ and [1$\bar{1}$0]$_\text{CrN}$ || [$\bar{1}\bar{1}$23]$_\text{AlN}$ (see Figure S4 in the supplementary material). For the CrN(111) layer, the full-width at half maximum (FWHM) of the CrN 111 x-ray rocking curve (XRC) is 0.1°, while for the CrN(113) layer, the FWHM values of the CrN 113 XRC are much larger: 2.2° along [1$\bar{1}$0]$_\text{CrN}$ and 3.5° along [33$\bar{2}$]$_\text{CrN}$.

The in-plane and out-of-plane lattice constants of the CrN(111) layer are calculated from a variety of symmetric, skew-symmetric, and asymmetric XRD reflections measured at room temperature, including 002, 111, 220, 222 and 313 (see Figure S5 in the supplementary material). The average lattice constant is determined to be (4.145 ± 0.002) Å, indicating minimal lattice distortion. This value is consistent with previously reported values for relaxed CrN.[3,11,14,15] This is plausible, given that the lattice mismatch between CrN(111) and AlN(0001) is approximately −5.8%, based on the calculated lattice constant. Such a substantial mismatch typically results in the rapid relaxation of the film, allowing CrN to quickly settle into a minimal strain state.

The residual strain of the 75-nm-thick CrN(111) layer is further analyzed using a Williamson-Hall plot (see Figure S6 in the supplementary material). The extracted microstrain is found to be 0.02%, indicating minimal internal stress. The crystallite size extracted from the Williamson-Hall analysis (using the Scherrer equation) ranges from 70 to 80 nm, which is consistent with the total film thickness. This suggests that this layer exhibits a high degree of crystallinity with large coherence lengths extending through the entire thickness. The consistency of crystallite size across symmetric, skew-symmetric, and asymmetric reflections indicates that the lateral and vertical grain dimensions are comparable, further suggesting a uniform and well-ordered structure. The relatively narrow XRC FWHM of 0.1° for the CrN(111) layer further supports a high crystalline quality and low defect density. These results strongly suggest that the CrN layer is fully relaxed, structurally uniform, and of high quality.

For the CrN(113) layer, its much larger orientational spread makes it challenging to measure multiple reflections for accurate lattice constant calculation, as done for the CrN(111) layer. However, due to the three-dimensional (3D) growth nature (see atomic force topographs in Figure S7 in the supplementary material) and asymmetric lattice mismatches along two orthogonal growth directions between CrN(113) and AlN(11$\bar{2}$2) (see surface arrangements in Figure S5(b) in the supplementary material), we expect the CrN(113) layer to be fully relaxed. Both CrN layers replicate the surface morphology of their AlN templates but exhibit 3D growth, which is attributed to N*-rich conditions and low-temperature growth.

For surface chemical analysis, x-ray photoelectron spectroscopy (XPS) measurements have been performed on the CrN(111) layer using a Scienta Omicron™ system equipped with an Al K$\alpha$ anode: $h\nu$ = 1486 eV) under ultra-high vacuum (UHV) conditions. XPS surveys are measured before and after Ar$^+$ sputtering with the intention to remove surface oxides and contaminants (see Figure S8 in the supplementary material). Ar$^+$ sputtering is applied multiple times until no further reduction in the intensities of the O 1$s$ and C 1$s$ peaks is observed. This suggests that while the surface has been largely cleaned, the remaining O 1$s$ and C 1$s$ peaks are likely due to O and C present in the layers (stemming from the source material) rather than surface contamination. Information regarding impurities in the Cr source, as provided by the commercial vendor, is presented in Figure S9 in the supplementary material. Additionally, the chemical composition of this clean layer is estimated to be Cr/N = 0.94 ± 0.06, indicating that the layer is nearly stoichiometric, which is consistent with the observed lattice constant and small residual strain. This is plausible as growth is performed under N*-rich conditions, which favor the formation of the CrN cubic rocksalt phase over trigonal Cr$_2$N.

To investigate the electrical properties of the CrN(111) and CrN(113) layers, Hall-effect measurements have been carried out using the van der Pauw configuration. Two setups were used to characterize two sets of CrN(111) and CrN(113) layers. The first one is a variable-temperature Hall setup with a cryostat enabling continuous temperature control between 4 and 400 K, and a magnetic field of 0.7 T. The second is a dedicated high-temperature setup (IPM vdW 1100K [66]) developed by the Fraunhofer Institut für Physikalische Messtechnik (IPM) enabling measurements up to 1100 K at a magnetic field of 0.5 T. A calibrated Pt-1000 resistance thermometer was used to measure the sample temperature with an accuracy of ±5°C. We used In contacts for the low-temperature measurements, and a direct mechanical contact of the PtRh sheathed thermocouples to the samples at high temperature. For the latter, the samples from set #2 were heated in an N$_2$ ambient to prevent oxidation and N desorption from the CrN surface. For the CrN(113) layer, contact issues prevented measurements above 750 K, while the CrN(111) layer remained stable up to 920 K. Between 300 and 400 K, both setups return almost identical resistivities and very similar electron densities and mobilities. None of the samples from set #2 exhibited any surface damage, indicating that the contact problems were technical and not physical in nature.

**Supporting Information**
Supporting Information is available from the Wiley Online Library.

**Acknowledgements**
This work is supported by the Deutsche Forschungsgemeinschaft (DFG, Germany Research Foundation) within the Priority Programme SPP2477 "Nitrides4Future – Novel Materials and Device Concepts" through Grant No. 563156864. The work of A.F. is funded by the BMFTR NanoMatFutur project All-GO-HEMT (03XP0630). We thank Carsten Stemmler for technical assistance with the MBE#9 system. The authors are grateful to Tauqir Shinwari for a critical reading of the manuscript, and to Markus Pristovsek at Nagoya University for supplying AlN templates.

**Conflict of Interest**
The authors declare no conflict of interest.

# Carrier transport and electrical bandgaps in epitaxial CrN layers:

# Supplementary material

*Duc V. Dinh,\* Jens Herfort, Andreas Fiedler, and Oliver Brandt*

Dr. Duc V. Dinh, Dr. Jens Herfort, Dr. Oliver Brandt
Paul-Drude-Institut für Festkörperelektronik, Leibniz-Institut im Forschungsverbund Berlin e.V., Hausvogteiplatz 5–7, 10117 Berlin, Germany.
E-mail: duc.vandinh@pdi-berlin.de
Dr. Andreas Fiedler
Leibniz-Institut für Kristallzüchtung, Max-Born-Straße 2, 12489 Berlin, Germany.

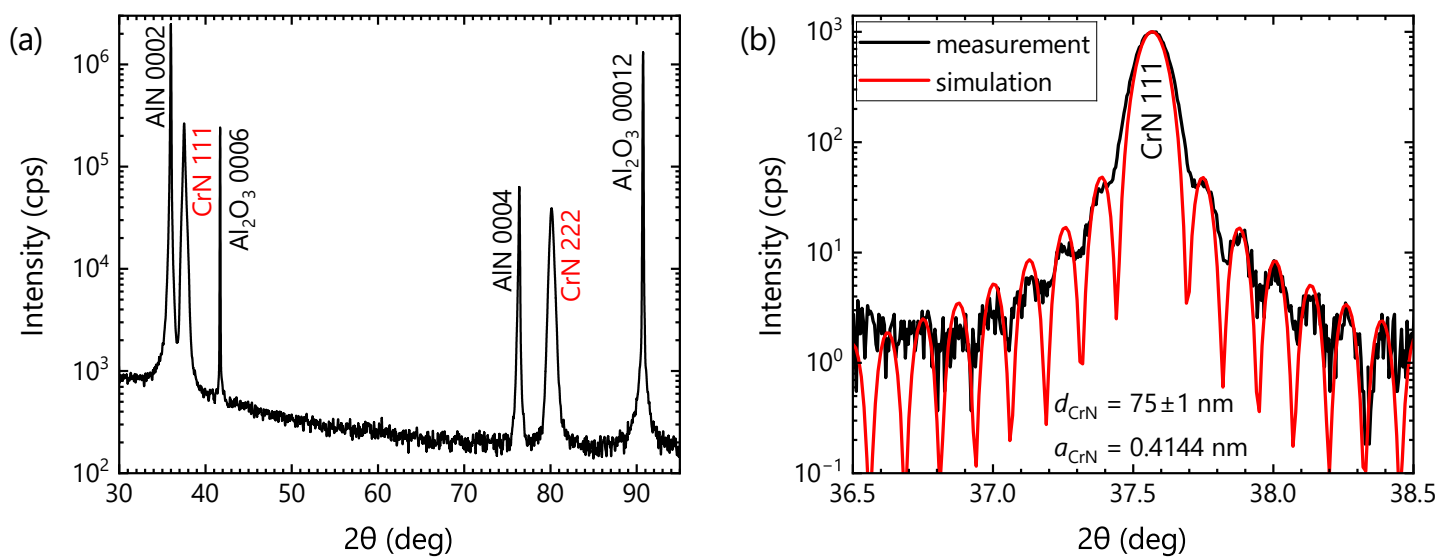

Figure S1: (a) Symmetric $2\theta - \omega$ x-ray diffraction (XRD) scan (performed with open detector, Philips PANalytical X'Pert PRO MRD) of the 75-nm-thick CrN(111) layer grown on an AlN(0001)/Al$_2$O$_3$(0001) template. (b) $2\theta - \omega$ XRD scan around the CrN 111 reflection (performed with analyzer) and corresponding simulation that indicates a thickness of 75 nm.

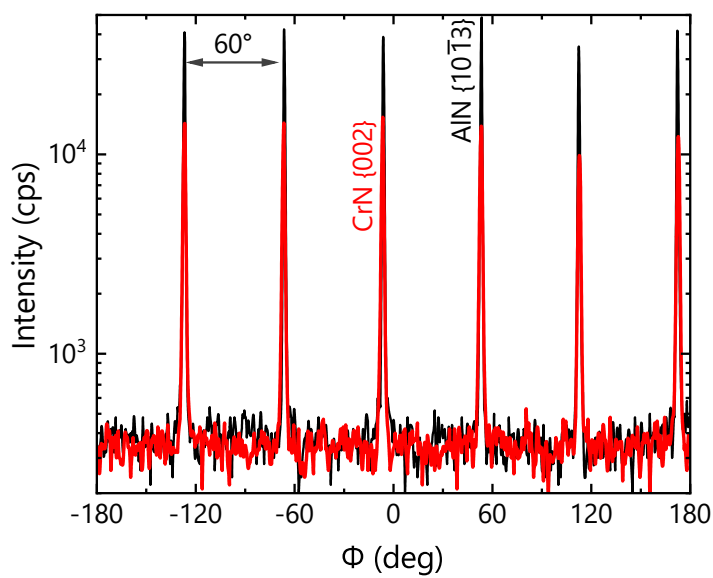

Figure S2: Azimuthal measurements of the CrN{002} ($\psi \approx 54.7°$) and AlN{10$\bar{1}$3} reflections ($\psi \approx 31.6°$). The six peaks indicates the twinned growth of CrN(111) on AlN(0001).

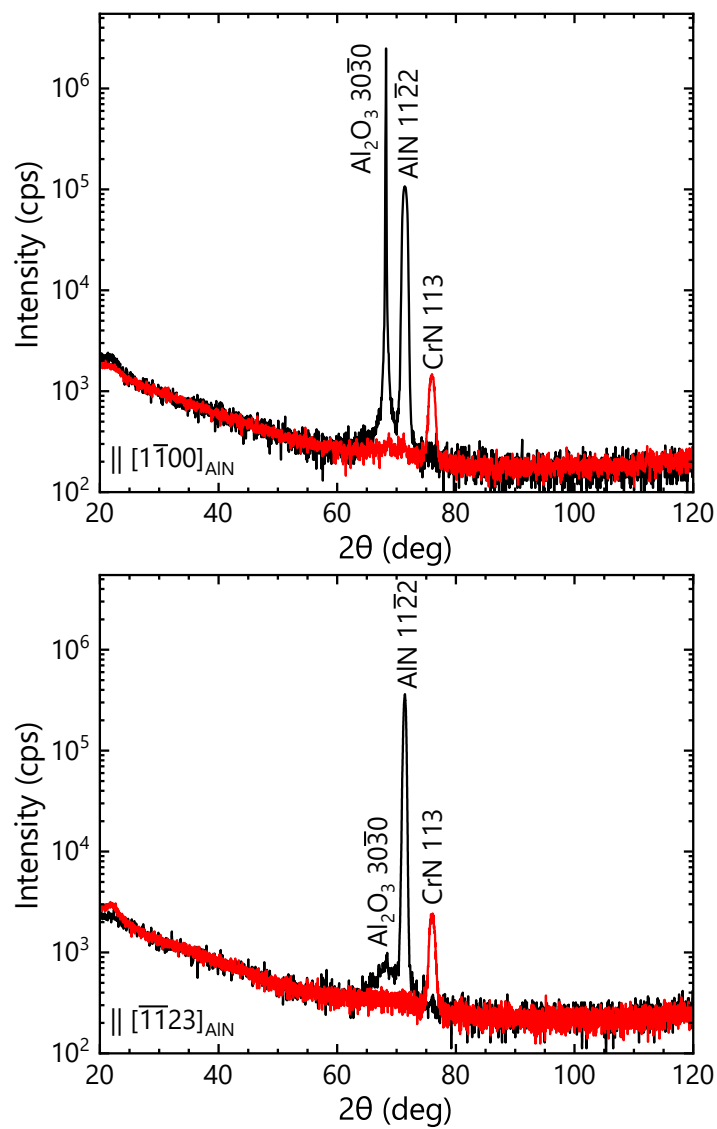

Figure S3: Symmetric $2\theta$–$\omega$ XRD measurements (performed with open detector) of the 75-nm-thick CrN layer grown on an AlN($11\bar{2}2$)/Al$_2$O$_3$($10\bar{1}0$) template scanned along [$1\bar{1}00$]$_{AlN}$ (top) and [$\bar{1}\bar{1}23$]$_{AlN}$ (bottom). Due to the significant lattice tilt between CrN(113) and AlN($11\bar{2}2$), the CrN113 and AlN$11\bar{2}2$ (Al$_2$O$_3$30$\bar{3}$0) reflections are not detected in a single scan, but require the sample to be precisely adjusted to the Bragg angle of the respective reflection.

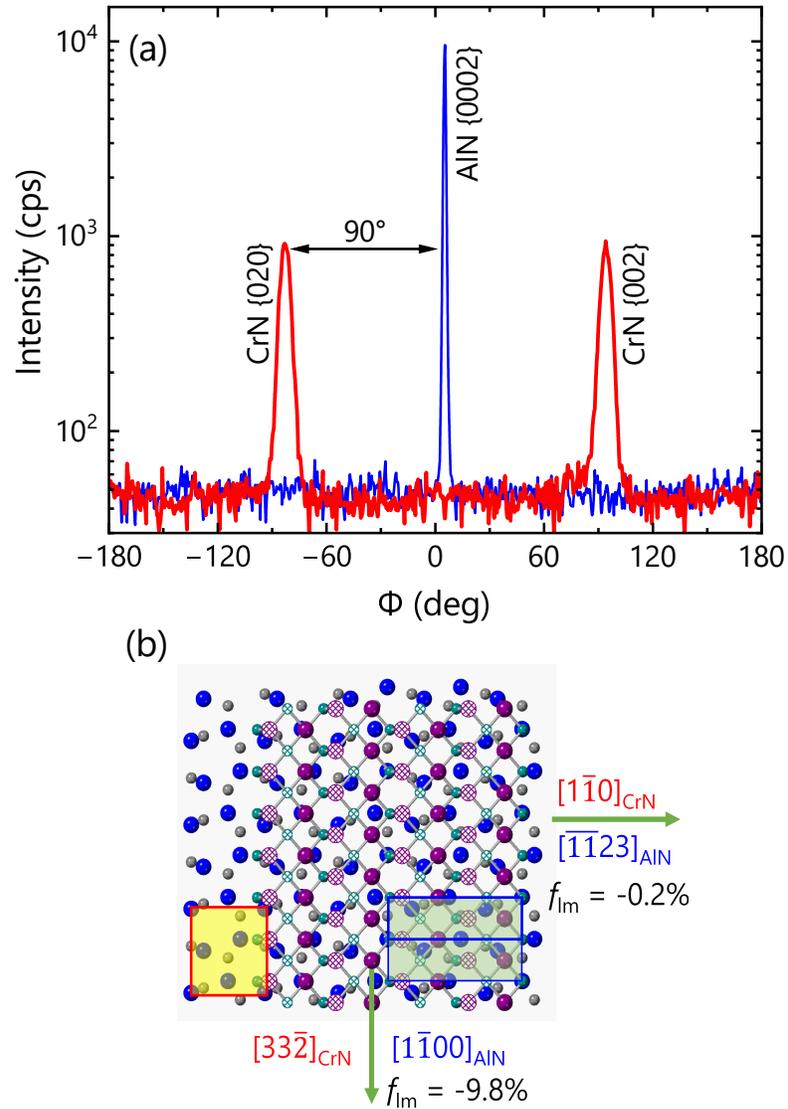

Figure S4: (a) Azimuthal measurements of the CrN{002} ($\psi \approx 25.2°$) and AlN{0002} reflections ($\psi \approx 58.0°$) indicates the twinned growth of the CrN(113)/AlN(11$\bar{2}$2) layer. (b) Surface atomic arrangements of CrN(113) on AlN(11$\bar{2}$2) indicate that 1 × 2 unit cells of CrN match to 2 × 1 unit cells of AlN. In (b), the epitaxial relationship between CrN and AlN, as well as estimated lattice-mismatch values ($f_{lm}$) along two in-plane directions are also shown.

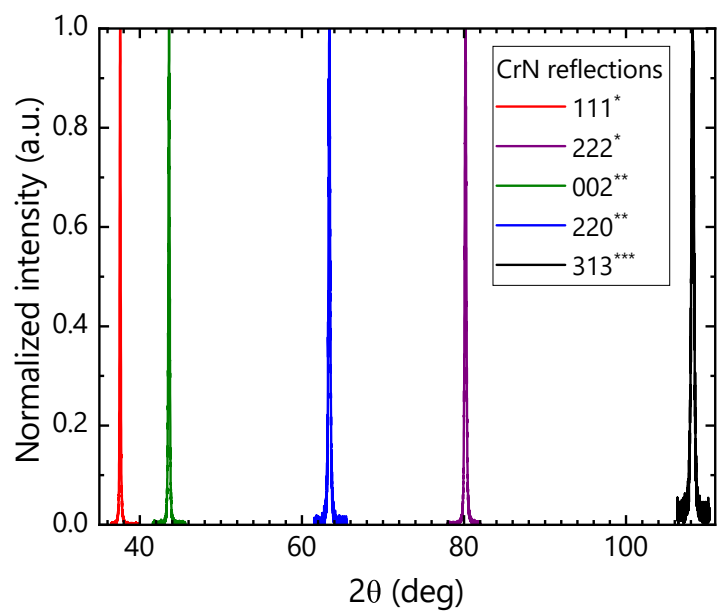

Figure S5: Symmetric (*), skew-symmetric (**), and asymmetric (***) XRD reflections measured at room temperature of the CrN(111) layer. From these measurements, the lattice constant of CrN is determined to be $(4.145 \pm 0.002)$ Å.

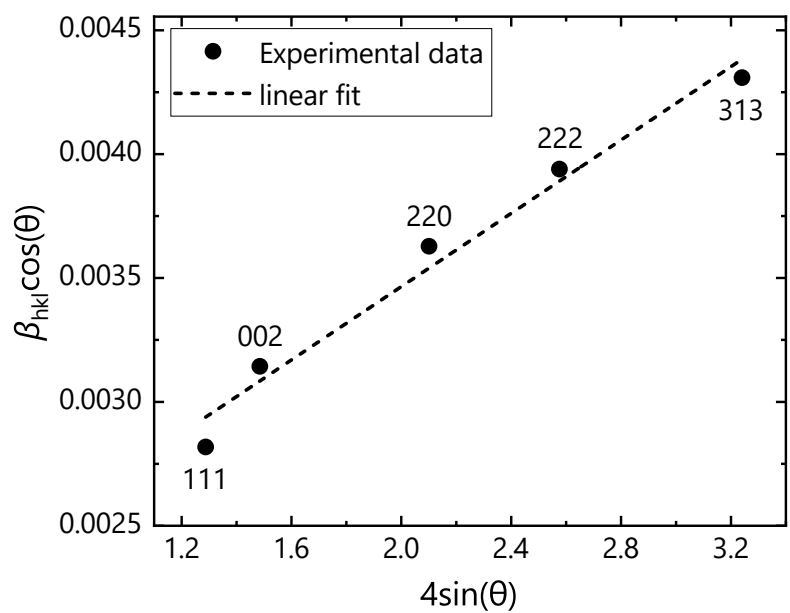

Figure S6: Williamson-Hall plot showing x-ray peak broadening ($\beta$) as function of Bragg angle ($\theta$) for the CrN(111) layer.

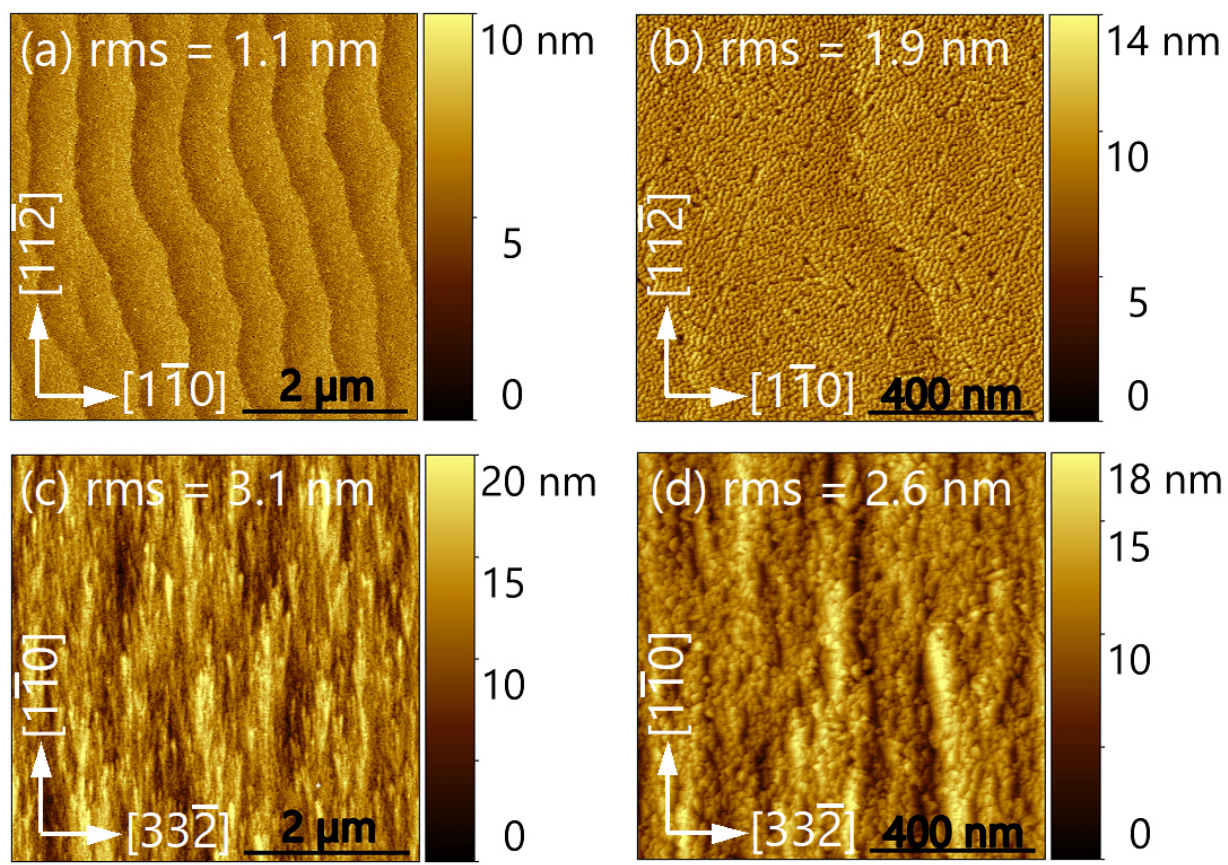

Figure S7: 5 × 5 and 1 × 1 µm² atomic force topographs (Dimension Edge, Bruker) of the CrN(111)/AlN(0001) layer (top row) and the CrN(113)/AlN(11$\bar{2}$2) layer (bottom row). Root-mean-square roughness (rms) values are shown for comparison.

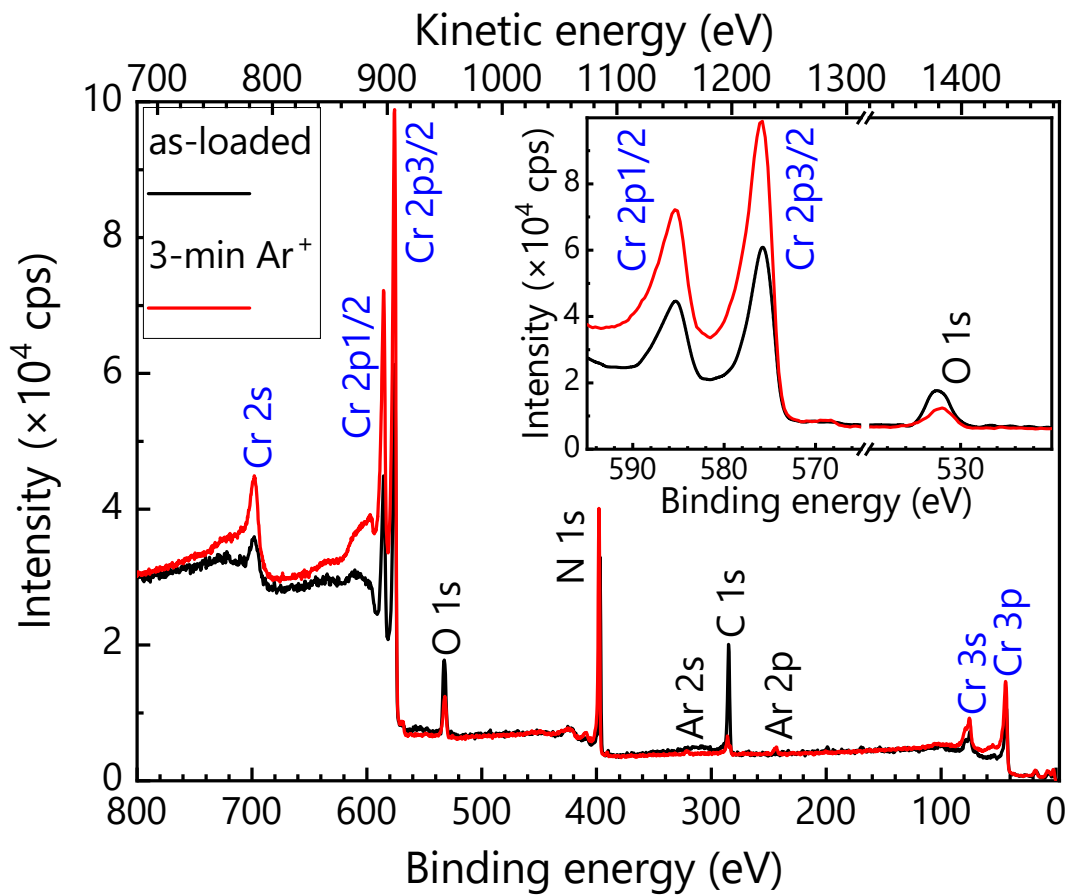

Figure S8: XPS survey spectra measured before and after Ar$^+$ sputtering (Scienta Omicron). Note the change in intensities of O 1$s$, Cr peaks, and C 1$s$ after Ar$^+$ sputtering. The insets show the spectra around Cr 2$p$ and O 1$s$ peaks.

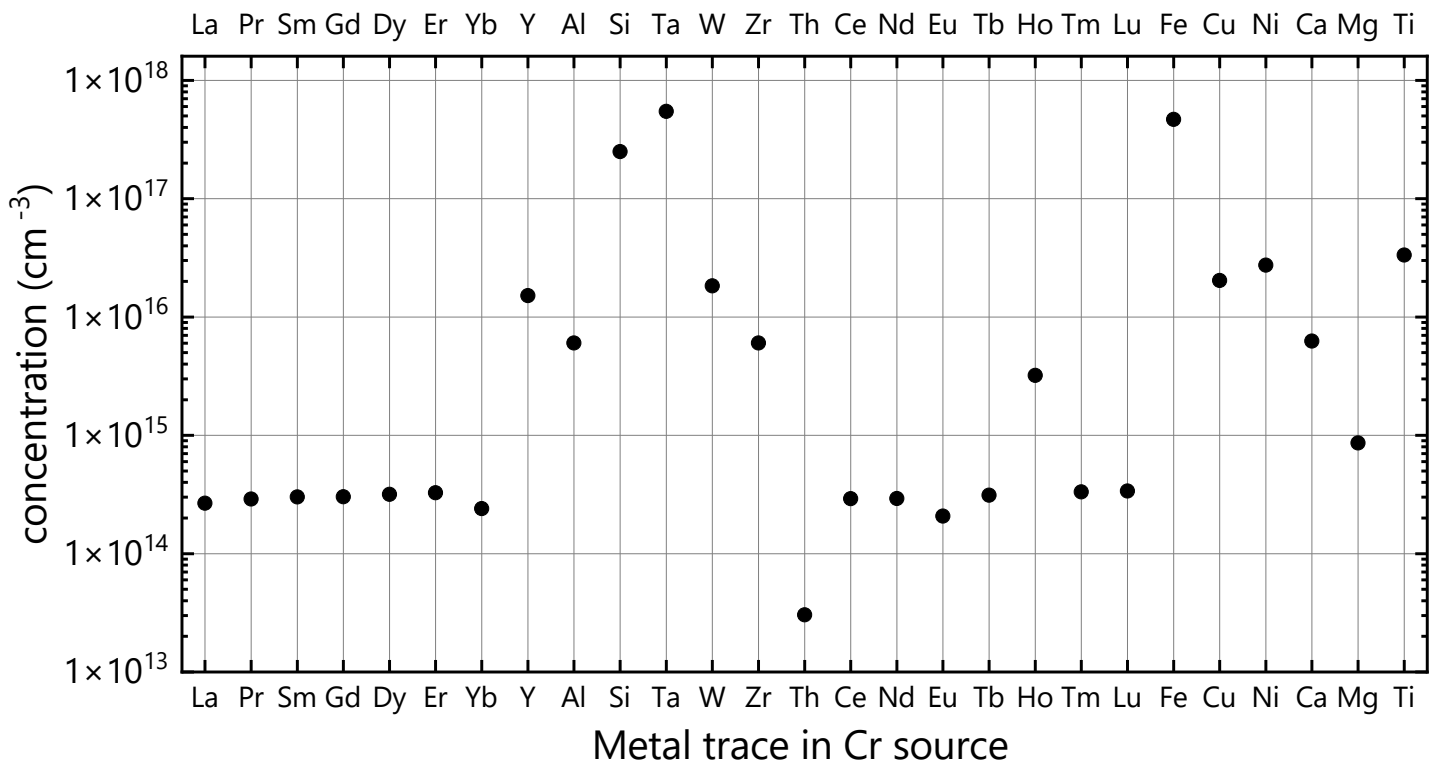

Figure S9: Metal and silicon impurity data in the Cr source (99.99% pure) as supplied by the commercial vendor.

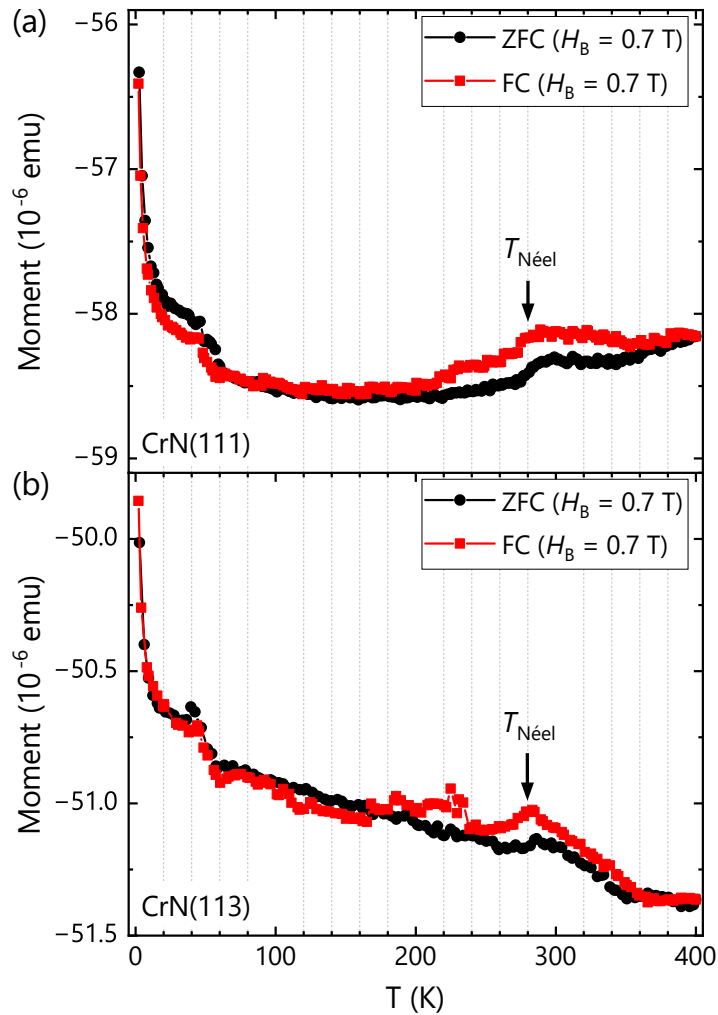

Figure S10: ZFC–FC measurements of (a) the CrN(111) and (b) CrN(113) layers from set #1.

The magnetic properties are studied using a commercial magnetic property measurement system (Quantum Design MPMS3) based on a superconducting quantum interference device (SQUID). Magnetization measurements were recorded at various temperatures with the magnetic field applied parallel to the sample surface. Before measuring the temperature dependence of the magnetization, the samples are demagnetized in an oscillating magnetic field at 300 K, then cooled to 2 K. Zero-field-cooled (ZFC) measurements are then performed from 2 K to 400 K under a saturation field of 0.7 T. After reaching 400 K, field-cooled (FC) measurements are continuously carried out from 400 K to 2 K under the same field. ZFC-FC measurements of both the CrN(111) and CrN(113) layers show clearly a $T_{\text{Néel}} \approx 280$ K, consistent with the value obtained for bulk antiferromagnetic CrN [1, 2, 3], and our electrical measurements (see the main text). Note that the small signal jumps observed at 40 K for both samples are likely due to instrumental artifacts rather than any intrinsic magnetic transition, and are currently under further investigation.

The slight divergence observed between the ZFC and FC magnetization curves below $T_{\text{Néel}}$ in the CrN films – despite the near-zero net magnetization expected for an ideal antiferromagnet – strongly suggests that domain-wall pinning plays a significant role. In the CrN(111) layer, this divergence is most pronounced between approximately 140 K and 280 K, where the splitting reaches up to $0.2 \times 10^{-6}$ emu at 280 K and diminishes to nearly zero around 200 K. While such a small splitting could also be influenced by the sensitivity limits of the SQUID, the systematic trend across temperature supports a physical origin linked to domain-wall dynamics. The twinned and rough microstructure of this layer introduces numerous pinning sites that hinder domain realignment during the ZFC process, resulting in lower magnetization compared to FC, where field cooling promotes better alignment. This divergence does not indicate ferromagnetic ordering but rather reflects extrinsic effects such as domain-wall pinning within an overall antiferromagnetic framework [4, 5]. Interestingly, below 140 K, both ZFC and FC curves in the CrN(111) layer increase similarly with decreasing temperature, suggesting a rise in magnetic susceptibility possibly linked to changes in antiferromagnetic order or increasing contributions from uncompensated surface spins as thermal fluctuations are further suppressed.

In contrast, the CrN(113) layer exhibits a minimal ZFC–FC divergence across the entire temperature range below $T_{\text{Néel}}$, despite its much rougher surface and poorer structural quality. This suggests that domain-wall pinning is already strong enough just below $T_{\text{Néel}}$ to freeze the domain configuration in both ZFC and FC processes, resulting in similar magnetization behavior. As temperature decreases further, this pinned configuration becomes more rigid, and both ZFC and FC signals increase gradually – much like the low-temperature behavior observed below 140 K in the CrN(111) layer. While additional effects such as spin canting, surface magnetism, or structural defects may also contribute to the non-ideal antiferromagnetic behavior, domain-wall pinning – closely tied to the layer's microstructure – emerges as the dominant mechanism behind the observed ZFC–FC trends.